\documentclass[%
preprint,
superscriptaddress,
groupedaddress,
showpacs,preprintnumbers,
 amsmath,amssymb,
 aps,
prc
]{revtex4-1}
\usepackage{graphicx}
\usepackage{dcolumn}
\usepackage{bm}

\usepackage{subfigure}
\usepackage{mathrsfs}
\usepackage{multirow}

\usepackage[colorlinks,
linkcolor={red!60!black!},
anchorcolor={green!80!black!},
urlcolor={blue!80!black!},
citecolor={blue!50!black!}]{hyperref}
\usepackage[table]{xcolor}

\begin{document}


\title{Nuclear multipole responses from chiral effective field theory interaction}

\author{B. S. Hu}
\author{Q. Wu}
\author{Q. Yuan}
\author{Y. Z. Ma}
\author{X. Q. Yan}
\author{F. R. Xu} \thanks{frxu@pku.edu.cn}
\affiliation{School of Physics, and State Key Laboratory of Nuclear Physics and Technology, Peking University, Beijing 100871, China}


\date{\today}

\begin{abstract}
We probe nuclear multipole resonances in the framework of the random-phase approximation by using the interaction obtained from the chiral effective field theory. The three-nucleon force is included in a form of the in-medium two-nucleon interaction which was derived from the chiral three-nucleon force. The isoscalar monopole, isoscalar dipole, isovector dipole and isoscalar quadrupole resonances of the closed-shell $^{56,68,78}$Ni have been investigated. The calculations reasonably reproduce the experimental multipole resonances of $^{56,68}$Ni, and well describe the pygmy dipole resonance and dipole polarizability  measured in $^{68}$Ni. The multipole resonances of $^{78}$Ni, including pygmy dipole resonance and dipole polarizability, are predicted. The detailed effects of the tensor force and three-body force are analyzed by dissecting the chiral interaction. We find that in general the tensor force effect on electric giant resonances is not as significant as the effect from the three-body force, although the tensor force provides more than half of the binding energy. The effect from three-body force is strong in light nuclei. Particularly, three-body force is crucial for the formation of the pygmy resonance in calculations.
\end{abstract}

\maketitle

\section{Introduction}
Giant resonances (GRs) in atomic nuclei are the most collective excitations
in which many nucleons participate in a joint motion with various multipolarities and different spin-isospin quantum numbers. 
GRs are relevant to many physics problems, ranging from finite nuclei to infinite nuclear matter to neutron stars and supernovae \cite{Glendenning1997,Harakeh2001}. The isoscalar giant monopole resonance (ISGMR) and isoscalar giant dipole resonance (ISGDR) provide direct way to probe the incompressibility of nuclei and nuclear matters \cite{Harakeh2001,GARG201855,Colo2014}. The electric dipole polarizability $\alpha_{\rm D}$ quantifies the behavior of dipole response and is related to the neutron distribution \cite{PhysRevC.81.051303,nphys3529}. 
The GRs of nuclei far from the valley of stability provide particular information on the structures of exotic nuclei and the equation of state (EOS) of neutron-rich matter. 

The ISGMR in neutron-rich nuclei can be used to probe the density dependence of the symmetry energy because it is sensitive to the incompressibility of nuclear matter \cite{PhysRevC.86.015802,PhysRevC.79.054311,Hebeler2014,Colo2014}. 
For neutron-rich nuclei, there exist low-lying electric dipole responses with weak strengths, named pygmy dipole resonance (PDR) \cite{CanJPhys,SAVRAN2013210}. They are interpreted as the dipole oscillations of excess neutrons against a core made by all other nucleons \cite{PhysRevC.3.1740}. The PDR can be used to determine the neutron-skin thickness and the parameters of the EOS. Neutron capture cross sections in the astrophysical $r$-process are also impacted by PDRs \cite{GORIELY199810,SAVRAN2013210}. 
Studying the evolution of GRs along an isotopic chain is useful in both experiment and theory for understanding of the isotopes and EOS. Nickel isotopes provide an excellent laboratory for the investigation of the evolution. 
In $^{56}$Ni, Monrozeau {\it et al.} \cite{PhysRevLett.100.042501} implemented the first measurement of the isoscalar response. Recently, the multipole response strengths in the neutron-rich $^{68}$Ni were observed \cite{PhysRevLett.111.242503,PhysRevLett.113.032504,PhysRevC.92.024316}. $^{78}$Ni with an extreme neutron-proton asymmetry (28 protons and 50 neutrons) is claimed to be a doubly magic nucleus \cite{nature569}. It was commented that there is a competition between spherical and deformed shapes, which is challenging the current theory \cite{nature569}.  

A variety of nonrelativistic and relativistic mean-field models have been successfully applied to the GRs of nuclei (see, e.g., \cite{Paar_2007,PhysRevLett.105.072501,VANGIAI19811,PhysRevC.96.031301,10.1143/PTPS.124.143,PhysRevLett.121.082501} and references therein). Calculations based on realistic nuclear forces have also made considerable progresses in the descriptions of the multipole responses, such as few-body approach \cite{PhysRevLett.96.112301,PhysRevC.91.024303}, no-core shell model \cite{QUAGLIONI2007370,Baker2019}, self-consistent Green's function (SCGF) \cite{PhysRevC.99.054327}, coupled cluster combined with the Lorentz integral transform \cite{PhysRevLett.111.122502,PhysRevC.94.034317} and Hartree-Fock plus random-phase approximation (HF-RPA) \cite{PhysRevC.74.014318,PhysRevC.75.014310,PhysRevC.97.054306}. 

In theory, the collective responses of nuclei are directly related to certain properties of the underlying nuclear force. The roles of the tensor force and three-nucleon force (3NF) have recently been highlighted in nuclear structure calculations. 
It has been claimed that the tensor force has a significant contribution to charge-exchange excitation \cite{BAI200928,PhysRevLett.105.072501,PhysRevLett.110.122501}. The 3NF has been known playing an important role in the first-principles calculations of nuclear matters and structures \cite{PhysRevC.86.054317,PhysRevC.91.051301,PhysRevLett.103.082501,PhysRevLett.106.202502,
PhysRevLett.110.242501,PhysRevLett.113.262504,RevModPhys.87.1067,PhysRevC.100.034324}. 
The chiral effective field theory (EFT) provides a robust framework to construct nucleon-nucleon interaction based on quantum chromodynamics \cite{Machleidt20111}. An important advantage of the chiral EFT is that it creates two- and three-nucleon forces on an equal footing. The chiral EFT interaction provides a good platform for analyzing the effects of the tensor force and 3NFs.

In the previous work \cite{PhysRevC.97.054306}, we calculated the monopole, dipole and quadrupole resonances of the closed-shell nuclei $^4$He,$^{16,22,24}$O and $^{40,48}$Ca using the chiral EFT NNLO$_{\rm sat}$ interaction within the Hartree-Fock plus RPA (HF-RPA) approach. The HF-RPA can reproduce experimental multipole resonances reasonably. In this work, we extend the calculations to heavier nuclei, $^{56,68,78}$Ni, with particular focus on the roles of the tensor force and 3NF of the underlying realistic nuclear force in the giant resonances.

\section{\label{sec:level2} The Hartree-Fock Random Phase Approximation (HF-RPA) }

The intrinsic Hamiltonian of the $A$-nucleon system reads
\begin{equation}
\label{eq1}
H=
\displaystyle\sum_{i=1}^{A} \left(1-\dfrac{1}{A}\right) \frac{\vec{p}_{i}^{2}}{2m} +
\displaystyle\sum_{i<j}^{A}   \left(V^{\rm NN}_{ij}-\frac{\vec{p}_{i}\cdot\vec{p}_{j} }{mA} \right) +\displaystyle\sum_{i<j<k}^{A} V^{\rm 3NF}_{ijk},
\end{equation}
where $V^{\rm NN}_{ij}$ is the two-body nucleon-nucleon ($\rm NN$) interaction,
and $V^{\rm 3NF}_{ijk}$ is the three-nucleon force (3NF).
The chiral EFT two-body interaction N$^3$LO developed by Entem and Machleidt \cite{Entem2003} is used. We include the 3NF via the in-medium two-body potential V$^{\rm 3NF_{\rm eff}}$ that was derived from the chiral N$^2$LO 3NF by integrating one nucleon over the Fermi sea (i.e., up to the Fermi momentum $k_{\rm F}$) in symmetric nuclear matter \cite{PhysRevC.79.054331,*PhysRevC.81.024002,PhysRevC.82.014314}. 
The extra low-energy constants for the  chiral effective N$^2$LO 3NF are $c_{\rm D}$ = $-$0.2 and $c_{\rm E}$ = 0.735 with the effective cutoff $\Lambda$ = 500 MeV and the Fermi momentum $k_{\rm F}$ = 0.95 fm$^{-1}$, which are the same as given in Ref.~\cite{PhysRevLett.109.032502}. With the chiral NN interaction at N$^3$LO \cite{Entem2003} and the effective in-medium 3NF$_{\rm eff}$ at N$^2$LO, the coupled cluster calculations have well described binding energies and low-lying excitation energies of heavy $pf$-shell nuclei \cite{PhysRevLett.109.032502}. 

The chiral NN+3NF$_{\rm eff}$ is expressed in 13 major harmonic oscillator (HO) shells with the commonly used oscillator frequency $\hbar \omega$ = 24 MeV \cite{PhysRevLett.109.032502,PhysRevC.99.061302,HENDERSON2018468}.
With the interaction established thus, we perform the HF-RPA calculations for the isoscalar monopole, isoscalar dipole, isovector dipole and isoscalar quadrupole resonances of the closed-shell nuclei $^{56,68,78}$Ni. The detail of the HF-RPA approach can be found in our previous article~\cite{PhysRevC.97.054306}.
The $E1$ photo-absorption cross section $\sigma(E)$ and electric dipole polarizability $\alpha_{D}$ are interesting observables for nuclear giant responses \cite{PhysRevC.99.054327,PhysRevC.94.034317}, which were not calculated in our previous paper on the He, O and Ca isotopes \cite{PhysRevC.97.054306}. The $\sigma(E)$ and $\alpha_D$ measure the responses to the isovector dipole resonance. We have 
\begin{equation}
\label{eq2}
\sigma(E)=4\pi^{2}\alpha E R(E),
\end{equation}
and
\begin{equation}
\label{eq3}
\alpha_{D}=2\alpha \int dE \dfrac{R(E)}{E},
\end{equation}
where $\alpha$ is fine-structure constant and $E$ is the excitation energy of the resonance. $R(E)$ is the response strength distribution of the $E1$ transition, 
\begin{equation}
R(E)=\sum_{\nu} B(E1,0\to\nu)\delta(E-\hbar\Omega_\nu),
\end{equation}
where $B(E1,0\to\nu)$ is the reduced electric dipole transition probability \cite{PhysRevC.97.054306}. The summation is over all possible particle-hole excitation modes \cite{PhysRevC.97.054306}. $\hbar\Omega_\nu$ stands for the excitation energy of an excitation mode which is obtained in the HF-RPA calculation \cite{PhysRevC.97.054306}. The discrete stength distributions $R(E)$ is smoothed by using the Lorentzian function,
\begin{equation}
R(E)=\sum_{\nu} B(E1,0\to\nu) \frac{1}{\pi} \frac{\Gamma/2}{(E-\hbar\Omega_{\nu})^2+(\Gamma/2)^2},
\end{equation}
where the width of $\Gamma$ = 2 MeV \cite{PhysRevC.74.014318,PhysRevC.75.014310,PhysRevC.97.054306} is used.

\section{\label{sec:results} Calculations and discussions}

\begin{figure*}
\setlength{\abovecaptionskip}{0pt}
\setlength{\belowcaptionskip}{0pt}
\includegraphics[scale=0.62]{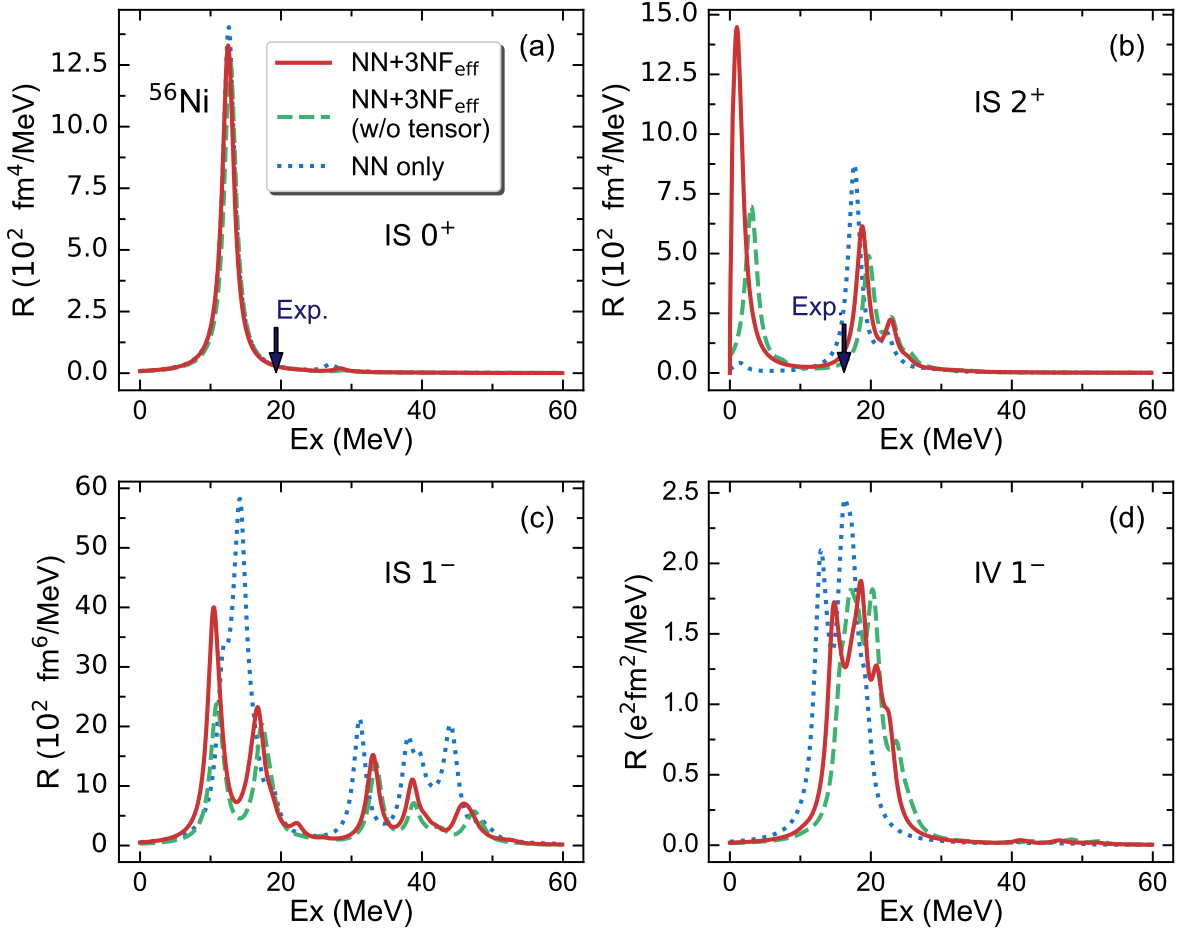}
\caption{\label{fig:Ni56} $^{56}$Ni isoscalar monopole (IS $0^+$), IS quadrupole (IS $2^+$), IS dipole (IS $1^-$) and isovector dipole (IV $1^-$) strength distributions calculated by HF-RPA. Different interactions are used: solid curves denote the calculations with the chiral N$^3$LO (NN) \cite{Entem2003} + in-medium effective 3NF (3NF$_{\rm eff}$) \cite{PhysRevC.79.054331,PhysRevC.81.024002,PhysRevC.82.014314}; dashed lines indicate the calculations by the chiral NN+3NF$_{\rm eff}$ but with the tensor terms being taken away; dotted lines label the calculations with N$^3$LO(NN) (tensor force included, but 3NF excluded). The experimental centroid energies \cite{PhysRevLett.100.042501} are indicated by arrows.}
\end{figure*}

Figure~\ref{fig:Ni56} gives the HF-RPA results for $^{56}$Ni, compared with available experimental data \cite{PhysRevLett.100.042501}. 
In the experiment \cite{PhysRevLett.100.042501}, the centroid energies of the ISGMR and isoscalar giant quadrupole resonance (ISGQR) were measured at 19.3$\pm$0.5 MeV and 16.2$\pm$0.5 MeV via the $^{56}$Ni($d$,$d^{\prime}$) reaction. The data provide useful information to test theoretical approaches and can be used to extract the nuclear matter incompressibility \cite{Harakeh2001,GARG201855,Colo2014}. 
In order to dissect the roles of the different components of the interaction, we use the spin-tensor decomposition method \cite{ELLIOTT1968241,KIRSON1973110,PhysRevC.74.034330} to decompose the interaction into central, tensor, and spin-orbit parts. The calculation without tensor force means that the tensor terms in NN+3NF$_{\rm eff}$ are taken away by using the spin-tensor decomposition method. 
As shown in Fig.~\ref{fig:Ni56}, the tensor force and 3NF have no significant effect on the IS monopole $0^+$ resonance, while their effects on other multipole resonances are seen clearly. The tensor force shifts the energy of the low-lying $2^+$ state by 2 MeV, and other IS quadrupole $2^+$ and IS dipole $1^-$ peaks are shifted within 5\% (the percent of the peak-energy shift over the peak energy). It changes the centroid energy of the IV dipole $1^-$ resonance by 16\%. It seems that the 3NF effects are more significant in the IS quadrupole $2^+$ and IS dipole $1^-$ resonances. 
The calculation without 3NF cannot clearly give the peak at $\sim$ 1 MeV in the IS quadrupole $2^+$ response distribution. 
The calculated RPA wave function shows that the low-lying $2^+$ state is caused by several one-particle one-hole excitations with the neutron or proton excited from $0f_{7/2}$ to $0f_{5/2}$ or $1p_{3/2}$. The peak seems to correspond to the experimental 2$^{+}_{1}$ excited state at $\sim$ 2.7 MeV \cite{nndc}. For the $2^+$ state, the present HF-RPA gives a reduced E2 transition probability $B(E2,0^+\to2^+)$ = 0.454 e$^2$b$^2$, larger than the experimental datum of 0.060(12) e$^2$b$^2$ \cite{PhysRevLett.73.1773}. The discrepancy between the calculations and data would originate from missing higher-order correlations in RPA \cite{PhysRevC.97.054306}.

\begin{figure*}
\centering
\setlength{\abovecaptionskip}{6pt}
\setlength{\belowcaptionskip}{6pt}
\includegraphics[scale=0.62]{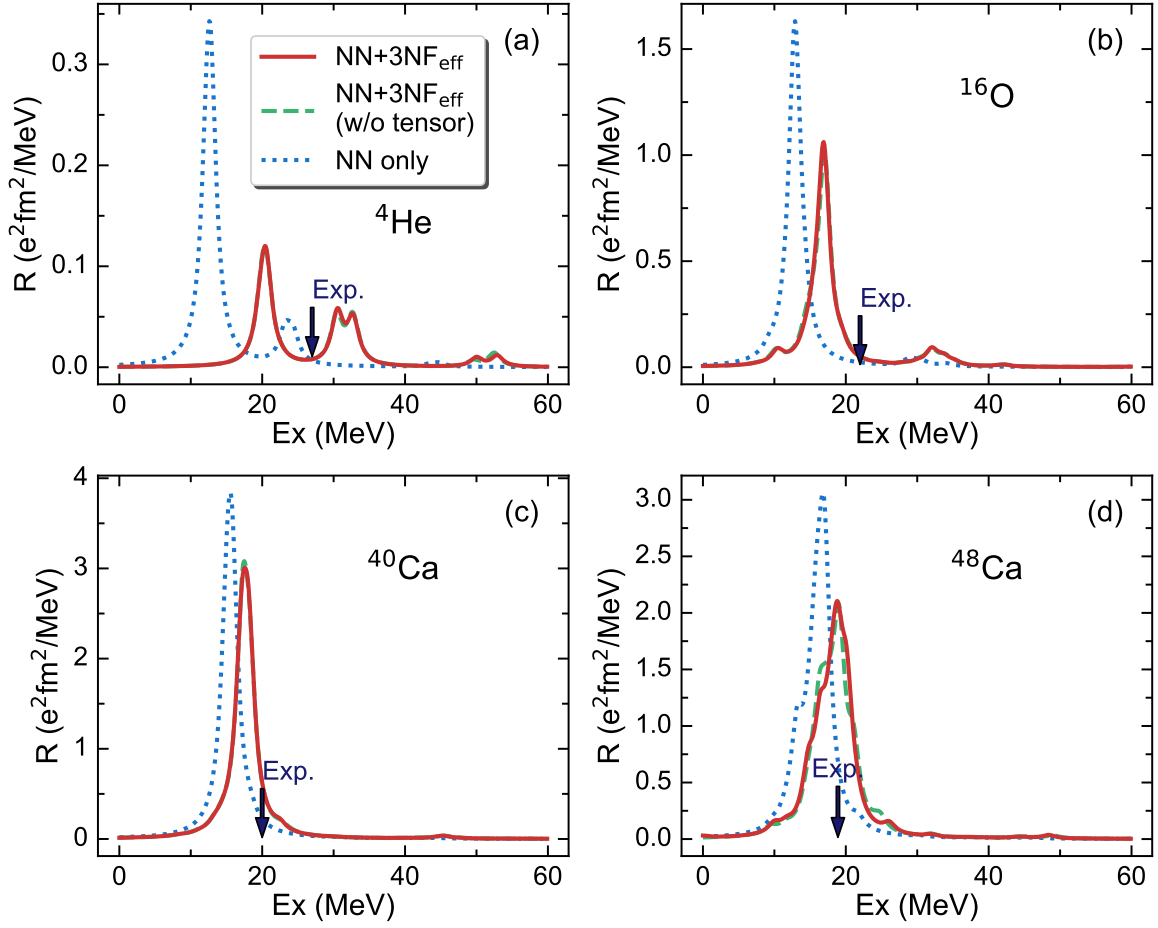}
\caption{Isovector dipole  (IV $1^-$) strength distributions in $^{4}$He, $^{16}$O and $^{40,48}$Ca. The experimental centroid energies are indicated by arrows, taken from Ref.~\cite{PhysRevLett.118.252501} for $^{48}$Ca and Ref.~\cite{RevModPhys.47.713} for other nuclei. The interactions used are same as in Fig. \ref{fig:Ni56}. }
\label{fig:IVD_compare}
\end{figure*}

We find that the 3NF effect on giant resonances is more pronounced in light nuclei. In Fig.~\ref{fig:IVD_compare}, we show the calculations of the IV dipole resonances for the closed-shell nuclei, $^4$He, $^{16}$O and $^{40,48}$Ca. The nuclei were investigated in our previous paper using the chiral NNLO$_{\rm sat}$ with the 3NF being normal ordered at a two-body level \cite{PhysRevC.97.054306}. Here, we recalculate their IV dipole resonances and analyze the 3NF effect by using the 3NF$_{\rm eff}$. Obtained results using the 3NF$_{\rm eff}$ are similar to those given by using NNLO$_{\rm sat}$ \cite{PhysRevC.97.054306}. 
We see that the 3NF effects are significant, particularly in $^4$He and $^{16}$O. 
Since we have no NNLO$_{\rm sat}$ interaction matrix elements for the heavy Ni isotopes, the N$^3$LO(NN) and NNLO(3NF$_{\rm eff}$) are used in the present calculations.

In Table~\ref{tab:alpha_D}, we calculate the dipole polarizability $\alpha_{\rm D}$ for the nuclei, compared with data \cite{PhysRevA.75.032521,AHRENS1975479,PhysRevLett.118.252501}, coupled-cluster (CC) \cite{PhysRevC.94.034317,nphys3529} and  density functional theory (DFT) \cite{COLO2013142} calculations. 
In the CC calculations \cite{PhysRevC.94.034317,nphys3529}, the chiral NNLO$_{\rm sat}$(NN+3NF) interaction was used. 
For the DFT results, the self-consistent RPA calculations with Skyrme forces SG$\rm \uppercase\expandafter{\romannumeral2}$~\cite{VANGIAI1981379}, SkM*~\cite{BARTEL198279}, SkP~\cite{DOBACZEWSKI1984103}, Sk255~\cite{PhysRevC.68.031304}, SLy4~\cite{CHABANAT1998231,*1998441}, Sly5~\cite{CHABANAT1998231,*1998441} and LNS~\cite{PhysRevC.73.014313} were performed by using the skyrme$_{-}$rpa code~\cite{COLO2013142}. The use of different Skyrme forces gives a range of the $\alpha_{\rm D}$ values, shown in Table~\ref{tab:alpha_D}.
The present and DFT calculations overestimate slightly the dipole polarizability $\alpha_{\rm D}$ in $^{4}$He, $^{16}$O and $^{48}$Ca, while in $^{40}$Ca the $\alpha_{\rm D}$ is slightly underestimated in the present and CC calculations. The tensor force has no significant effect on the electric dipole polarizability $\alpha_{\rm D}$, as shown in Table~\ref{tab:alpha_D}. 
\newcommand{\tabincell}[2]{\begin{tabular}{@{}#1@{}}#2\end{tabular}}
\begin{table}
\caption{ \label{tab:alpha_D}%
Calculated isovector dipole polarizability $\alpha_{\rm D}$ with and without 3NF (and with or without the tensor force), compared with data \cite{PhysRevA.75.032521,AHRENS1975479,PhysRevLett.118.252501}, coupled cluster (CC) \cite{PhysRevC.94.034317,nphys3529} and density functional theory (DFT) \cite{COLO2013142} calculations. N$^3$LO(NN) \cite{Entem2003} and N$^2$LO(3NF$_{\rm eff}$) \cite{PhysRevC.79.054331,PhysRevC.81.024002,PhysRevC.82.014314,PhysRevLett.109.032502} are used. See the text for the DFT calculations.}
\begin{tabular*}{150mm}{l@{\extracolsep{\fill}}ccccccc p{cm}}
\toprule 
 & \tabincell{c}{NN only \\ (with tensor)} & \tabincell{c}{NN+3NF$_{\rm eff}$ \\ (w/o tensor)} & \tabincell{c}{NN+3NF$_{\rm eff}$ \\ (with tensor)} & Exp. & CC & DFT\\
\hline
$^{4}$He & 0.2811 & 0.0892 & 0.0893 & 0.076(8) & 0.0735 & 0.1108$-$0.1333\\
$^{16}$O & 1.2714 & 0.7618 & 0.7593 & 0.58(1) & 0.58 & 0.6154$-$0.6922\\
$^{40}$Ca & 2.5866 & 2.0521 & 2.0507 & 2.23(3) & 2.08 & 2.0175$-$2.2508\\
$^{48}$Ca  & 2.8391 & 2.2231 & 2.2438 & 2.07(22) & 2.19$-$2.60 & 2.3594$-$2.5813\\
\hline
\hline
\end{tabular*}
\end{table}

\begin{figure*}
\setlength{\abovecaptionskip}{0pt}
\setlength{\belowcaptionskip}{0pt}
\includegraphics[scale=0.62]{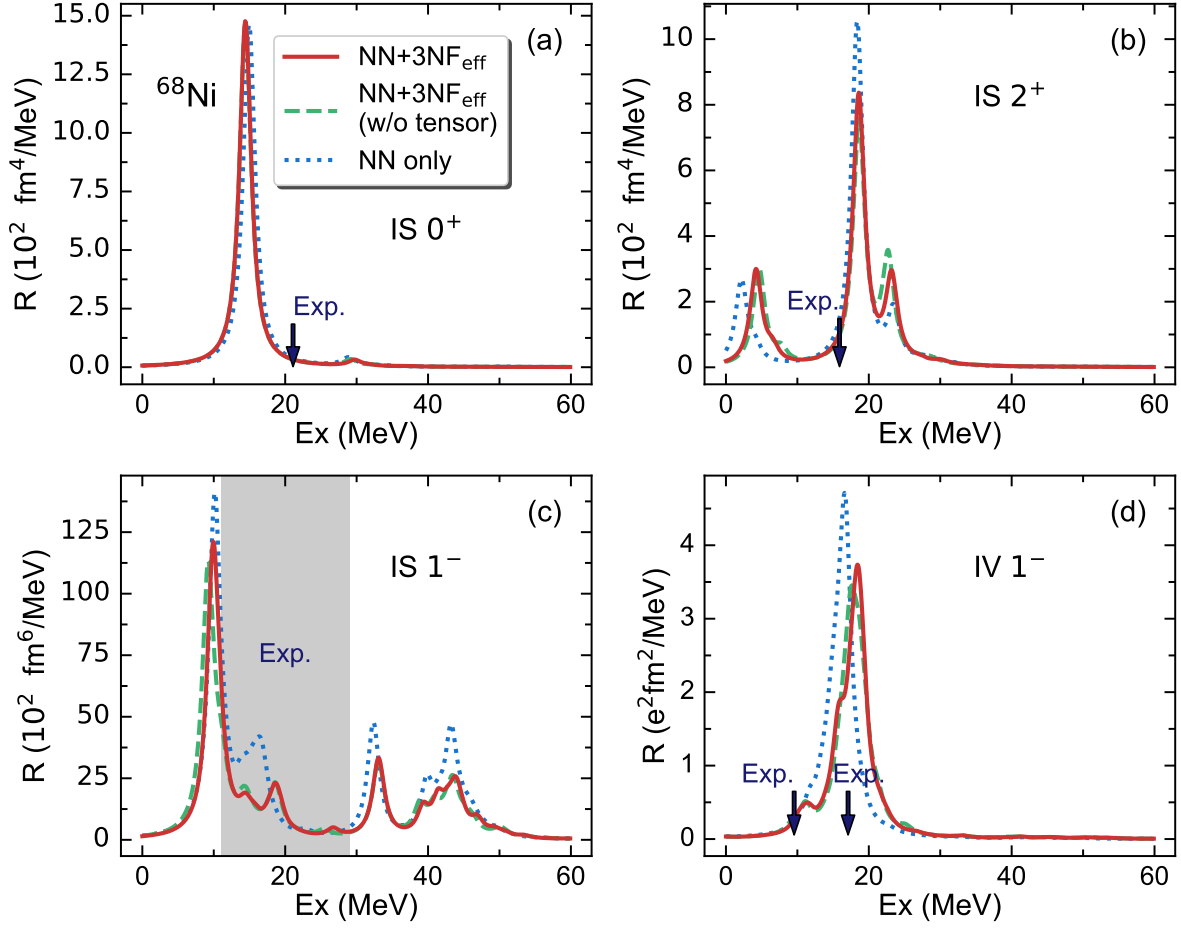}
\caption{\label{fig:Ni68} Similar to Fig.~\ref{fig:Ni56} but for $^{68}$Ni. The experimental centroid energies \cite{PhysRevLett.111.242503,PhysRevLett.113.032504,PhysRevC.92.024316} are indicated by arrows. The gray shadowing in panel (c) indicates a possible IS dipole resonance in the energy range of 11$-$29 MeV given by the experiment \cite{PhysRevC.92.024316}.}
\end{figure*}

Figure~\ref{fig:Ni68} shows the results for $^{68}$Ni. It is seen that the effect of tensor force is similar to that in $^{58}$Ni. In the neutron-rich $^{68}$Ni, the isovector PDR (IVPDR) and isovector GDR (IVGDR) were observed recently by Rossi {\it et al.} \cite{PhysRevLett.111.242503} with peaks located at 9.55(17) MeV and 17.1(2) MeV, respectively. We see that the calculation with 3NF can give the IV $1^-$ PDR peaked at energy $\approx$ 10 MeV,
while the calculation without 3NF does not show a clear PDR peak. However, the reduced electric dipole transition probability $B(E1)$ calculated without 3NF shows a possible weak PDR at energy $\approx$ 10 MeV. There is a strong resonance peak at $\approx$ 10 MeV in the IS $1^-$ channel, which overlaps with the weak IV $1^-$ peak. This indicates a mixing nature of isoscalar and isovector resonances in the dipole channel at low energy. The mixing was seen in our previous calculations \cite{PhysRevC.97.054306} for the neutron-rich $^{22,24}$O. Experimentally, the IS dipole resonance can be incurred by inelastic scatterings with an isoscalar particle (e.g., $\alpha$ particle), while electromagnetic excitations (usually by electron scattering) give the total strength of the IS and IV resonances. The recent experiment \cite{PhysRevC.92.024316} has shown the possible IS dipole resonances in the energy range of $\approx$ 11$-$29 MeV (see Fig.~\ref{fig:Ni68}). 

The experiment by Rossi {\it et al.} \cite{PhysRevLett.111.242503} presents the first measurement of the dipole polarizability $\alpha_{\text D}$ in an unstable neutron-rich nucleus. Using the measured $\alpha_{D}$, the authors \cite{PhysRevLett.111.242503} deduced a neutron-skin thickness of 0.17(2) fm in $^{68}$Ni by taking a nearly linear relation between $\alpha_{D}$ and neutron-skin thickness guided by the relativistic RPA calculation \cite{PhysRevC.83.034319}. The data provide further constraint on the isospin-asymmetric part of the EOS.   

Table~\ref{tab:IVGDR} lists the calculated isovector dipole polarizability, compared with the data \cite{PhysRevLett.111.242503}, SCGF \cite{PhysRevC.99.054327} and DFT \cite{COLO2013142} calculations. The calculated dipole polarizability agrees well with the data \cite{PhysRevLett.111.242503} and SCGF \cite{PhysRevC.99.054327}. 
In the SCGF calculations, NNLO$_{\rm sat}$(NN+3NF) was used.
There are two IV dipole resonance peaks at the low energies of  10.6 and 11.6 MeV, similar to the SCGF conclusions.
The mean-field calculation \cite{PhysRevC.84.051301} predicted a possible soft monopole excitation in $^{68}$Ni. However, we do not see the soft monopole mode in the present calculations, while there is a low-lying IS quadrupole response peak at the energy of $\approx$ 4.2 MeV. 
The low-energy IS quadrupole peak would correspond to a soft resonance involving few-particle few-hole excitations or even one-particle one-hole excitation as happening in $^{22,24}$O \cite{PhysRevC.97.054306}. These may need to be verified further by experiments and other models. In Table~\ref{tab:IVGDR}, we also show the results of $^{78}$Ni as prediction.
\begin{table}
\caption{ \label{tab:IVGDR}%
Calculated excitation energies of isovector PDR and GDR in $^{68,78}$Ni, compared with the experimental data \cite{PhysRevLett.111.242503,PhysRevLett.102.092502}, SCGF \cite{PhysRevC.99.054327} and density functional theory (DFT) \cite{COLO2013142} results. The chiral N$^3$LO(NN) \cite{Entem2003}+in-medium effective 3NF (3NF$_{\rm eff}$) \cite{PhysRevC.79.054331,PhysRevC.81.024002,PhysRevC.82.014314} are used. $S_{\rm PDR}$ indicates the percentage of the energy-weighted sum rule (for the isovector PDR) with respect to the classical Thomas-Reiche-Kuhn (TRK) value.  The $S_{\rm PDR}$ is obtained by summing the strength distributions of the PDRs up to 12 MeV in the calculations.  $\alpha_{\text D}$ is the dipole polarizability of the isovector channel. The DFT calculations are similar to Table~\ref{tab:alpha_D}.}
\begin{tabular*}{150mm}{l@{\extracolsep{\fill}}cccccc}
\hline
\hline  
\multicolumn{1}{c}{} & \multicolumn{4}{c}{$^{68}$Ni} & \multicolumn{2}{c}{$^{78}$Ni} \\
\cline{2-5}
\cline{6-7}
& Exp.  & Present & SCGF & DFT & Present & DFT\\
\hline
\multirow{2}{*}{$E_{\rm PDR}$ (MeV)} & \multirow{2}{*}{9.55(17)} & 10.64 & 10.68 &\multirow{2}{*}{8.41$-$10.60} & \multirow{2}{*}{11.03} & \multirow{2}{*}{8.78$-$11.42}\\
  &  & 11.55 & 10.92 & \\
$S_{\rm PDR}$ (\%) & 2.8(5) \cite{PhysRevLett.111.242503}; 5 \cite{PhysRevLett.102.092502}  & 5.20  & $-$ & 1.77$-$3.56 & 5.81 & 2.00$-$5.59\\
$E_{\rm GDR}$ (MeV) & 17.1(2) & 18.64 & 18.10 & 16.44$-$17.95 &17.17 & 15.71$-$18.72\\
$\alpha_{D}$ (fm$^3$)  & 3.40(23) & 3.40& 3.60 & 3.99$-$4.52& 3.76 & 4.48$-$5.26\\
\hline
\hline
\end{tabular*}
\end{table}

The sum rule of the response strength distribution can be used to analyze the interaction in the momentum dependence and isospin exchange \cite{Erler_2011}. The energy-weighted sum rule (EWSR) is defined by \cite{Orlandini_1991} 
\begin{eqnarray}
S(E1)=\sum\limits_\nu \hbar\Omega_\nu B(E1,0 \rightarrow \nu)=\dfrac{\hbar^2e^2}{2m} \dfrac{9}{4\pi}\dfrac{NZ}{A}(1+\kappa),
\end{eqnarray}
where $\kappa$ is the so-called enhancement factor which can be obtained by integrating the strength function \cite{PhysRevC.97.054306}. As shown in Table~\ref{tab:IVGDR}, the present calculation with 3NF$_{\rm eff}$ gives that, in $^{68}$Ni, the strength below 12 MeV exhausts 5.2\% of the classical Thomas-Reiche-Kuhn (TRK) energy-weighted sum rule (EWSR), compared with the experimentally extracted value of $\approx 5\%$ \cite{PhysRevLett.102.092502} or 2.8(5)\% \cite{PhysRevLett.111.242503}. Note that the PDR peak at 10.64 MeV contributes about 2.0\% of the EWSR, and the 11.55 MeV peak contributes about 3.2\%. We predict that it is about 5.8\% in $^{78}$Ni. 
Table~\ref{tab:IVGDR} also gives the DFT results, as discussed in Table~\ref{tab:alpha_D}. The DFT calculations give smaller EWSR but larger $\alpha_{\rm D}$ than the present calculations in $^{68,78}$Ni.

\begin{figure*}
\setlength{\abovecaptionskip}{0pt}
\setlength{\belowcaptionskip}{0pt}
\includegraphics[scale=0.62]{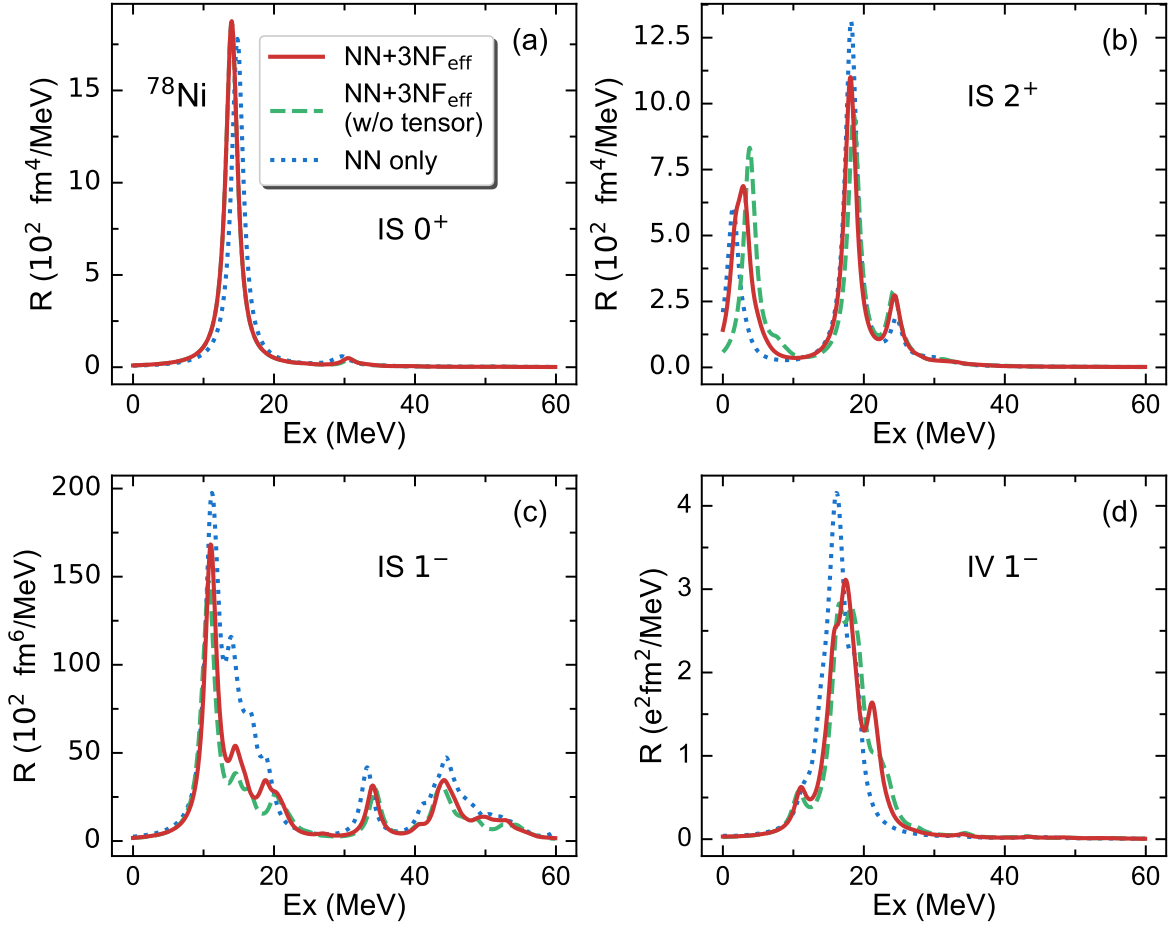}
\caption{\label{fig:Ni78} Similar to Figs.~\ref{fig:Ni56} and \ref{fig:Ni68}, but for $^{78}$Ni.}
\end{figure*}

Figure~\ref{fig:Ni78} displays the calculated ISGMR, ISGQR, ISGDR, and IVGDR for $^{78}$Ni, predicting a IV $1^-$ PDR peaked at 11 MeV.
We see that the PDR is enhanced in  the calculation with 3NF.
The calculated centroid energies of the isovector PDR and GDR are given in Table~\ref{tab:IVGDR}. 
The isovector GDR energy in $^{78}$Ni is 1.5 MeV lower than the one in $^{68}$Ni. The dipole polarizability $\alpha_{\rm D}$ is also predicted in Table~\ref{tab:IVGDR}. $^{78}$Ni has been believed to be a doubly magic nucleus, with a high $2^+_1$ excitation energy at 2.6 MeV \cite{nature569}. The present HF-RPA calculation with 3NF gives 2.90 MeV for the $2_1^+$ excited state, which is consistent with the {\it ab-initio} coupled cluster and in-medium similarity renormalization group calculations given in Ref. \cite{nature569}.

\begin{figure}
\setlength{\abovecaptionskip}{0pt}
\setlength{\belowcaptionskip}{0pt}
\includegraphics[scale=0.62]{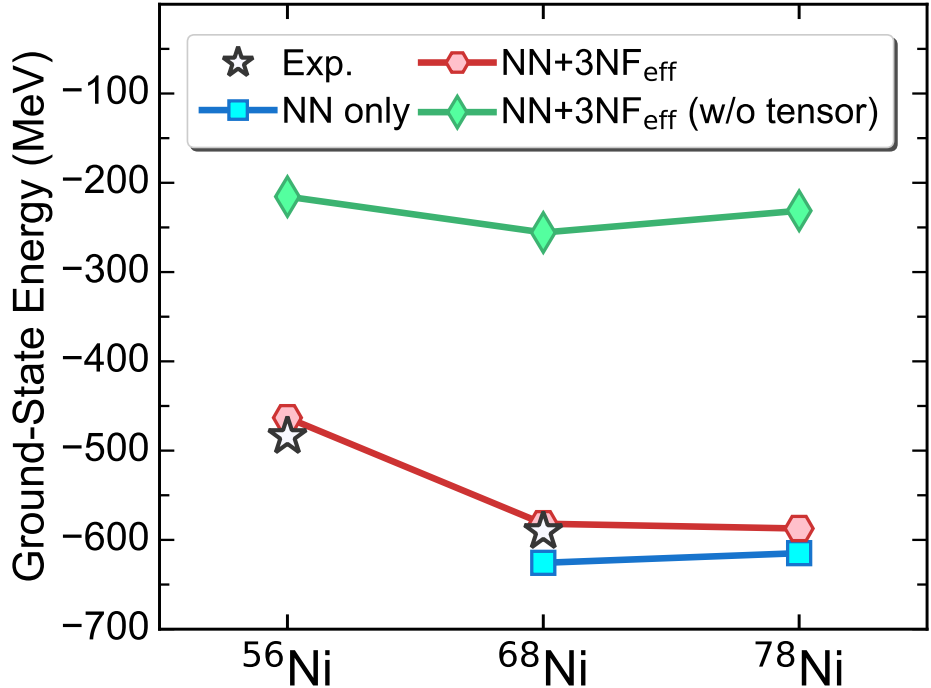}
\caption{\label{fig:BindingEnergy} $^{56,68,78}$Ni ground-state energies calculated by HF-RPA with the second-order perturbation correction included.}
\end{figure}

From the calculations discussed above, we see that 3NF effects on collective multipole resonances are meaningful, particularly for light nuclei. The 3NF (including tensor ingredients)  plays a crucial role in the formations of the PDRs in $^{68, 78}$Ni. 
Fig.~\ref{fig:BindingEnergy} plots the binding energies calculated within the RPA framework.
In the binding energy calculation, the correction from the second-order perturbation \cite{PhysRevC.97.054306,PhysRevC.94.014303} is considered.
We see that the tensor force provides more than 50\% of the binding energy. If tensor components were taken away from the realistic interaction, we should not be able to describe the ground states of nuclei correctly. The 3NF also has a significant effect on binding energy, and improves the calculation, see Fig.~\ref{fig:BindingEnergy}. We find that the second-order many-body perturbation correction is not converged in $^{56}$Ni if only two-body interaction is considered in the calculation. This is why in Fig.~\ref{fig:BindingEnergy} the $^{56}$Ni binding energy is missing in the NN-only cure. 

In realistic nuclear forces, the tensor component is large. For example, in the chiral EFT leading-order term that consists of one-pion exchange and contact interactions, the tensor force has the same strength as the central force in the one-pion exchange \cite{Machleidt20111,Tanihata_2013,SAGAWA201476}. The leading-order term should be most important for the calculations of binding energy and other observables. The result that the tensor force provides a large proportion of the binding energy is quite general in calculations based on realistic forces. In Refs.~\cite{Tanihata_2013,10.1143/PTPS.56.32}, the calculation with the Hamada-Johnston and Tamagaki interaction shows that about 50\% of the $^4$He potential energy comes from the tensor force. In the Green-function Monte-Carlo calculations \cite{annurev.nucl.51.101701.132506}, the one-pion exchange in the AV18 interaction provides 70\%-80\% of the nuclear potential energy for light nuclei. With the present method and interaction, we have also calculated the binding energies of $^4$He, $^{16}$O and $^{40,48}$Ca, giving the similar result, i.e., the tensor force provides about 50\% of the binding energy.

\section{\label{sec:summary and outlook} Summary}
Starting from the chiral effective field theory interaction N$^3$LO(NN)+N$^2$LO(3NF$_{\rm eff}$), we have performed the RPA calculations within the Hartree-Fock (HF) approach, to investigate the monopole, dipole and quadrupole resonances in $^{56,68,78}$Ni. The ground-state energies have also been calculated by incorporating the second-order perturbation correction into the HF-RPA energy. The present HF-RPA calculations reproduce reasonably the multipole resonances observed  in $^{56, 68}$Ni and their binding energies as well. The pygmy dipole resonance, the dipole polarizability $\alpha_{\rm D}$ and the sum rule in $^{68}$Ni have been discussed, and compared with experimental data available. The properties of $^{78}$Ni have been predicted.

We dissect the 3NF and tensor terms of the realistic interaction, to see their roles in the multipole resonances of nuclei. Although the tensor force may be important for charge-exchanged collective resonances, it is not such significant for the electric giant resonances. However, the tensor force provides more than half of binding energy using the chiral EFT interaction. The three-body force has a nonnegligible effect on multipole resonances, particularly on the formation of the pygmy resonance. The tensor force and three-body force hardly affect the isoscalar monopole giant resonance. In possible future work, we will extend the present framework to charge-exchange excitations.

\begin{acknowledgments}
We thank Umesh Garg for his useful suggestions and discussions.
This work has been supported by
the National Key R${\&}$D Program of China under Grant No. 2018YFA0404401;
the National Natural Science Foundation of China under Grants No. 11921006, No. 11835001 and No. 11847203;
China Postdoctoral Science Foundation under Grant No. 2018M630018;
and the CUSTIPEN (China-U.S. Theory Institute for Physics with Exotic Nuclei) funded by the U.S.  Department of Energy,
Office of Science under Grant No. DE-SC0009971.
We acknowledge the High-performance Computing Platform of Peking University for providing computational resources.
\end{acknowledgments}

\bibliography{references}

\end{document}